# Introduction to Artificial Consciousness: History, Current Trends and Ethical Challenges

Aïda Elamrani

**Abstract**

With the significant progress of artificial intelligence (AI) and consciousness science, artificial consciousness (AC) has recently gained popularity. This work provides a broad overview of the main topics and current trends in AC. The first part traces the history of this interdisciplinary field to establish context and clarify key terminology, including the distinction between Weak and Strong AC. The second part examines major trends in AC implementations, emphasising the synergy between Global Workspace and Attention Schema, as well as the problem of evaluating the internal states of artificial systems. The third part analyses the ethical dimension of AC development, revealing both critical risks and transformative opportunities. The last part offers recommendations to guide AC research responsibly, and outlines the limitations of this study as well as avenues for future research. The main conclusion is that while AC appears both indispensable and inevitable for scientific progress, serious efforts are required to address the far-reaching impact of this innovative research path.

# 1 Introduction

Over recent years, Artificial Consciousness (AC) has gained significant momentum. Within the AI community, computational models of consciousness are increasingly explored as frameworks for enhancing artificial agents' capabilities (Bengio, 2019; Chella, 2023). A simple Google Scholar search reflects this growing interest: between 1950 and 2009, only 166 results appear for "artificial consciousness" or "machine consciousness"; this number rises to 343 for the period 2010–2019 and surges to 480 since 2020.

Meanwhile, rapid progress in AI and the proliferation of novel commercial applications have blurred the lines in how experts and the general public perceive AI. Surveys reveal that many people believe current AI systems may already possess some form of consciousness, and a majority expect artificial systems to become sentient within the next 100 years (Pauketat et al., 2022, 2023; Scott et al., 2023; Colombatto & Fleming, 2024). These beliefs are accompanied by growing ethical concerns about the moral status of digital artefacts (Finlinson, 2025; Long & Sebo, 2024; Shevlin, 2024b; AMCS, 2023).

Our aim here is to provide a broad overview of AC that is accessible to novices yet insightful to experts. This overview is structured around three key axes.

First, we define what AC is by retracing the field's history and how the convergence of AI and the scientific study of the human mind gave rise to the study of artificial consciousness. We analyse key philosophical debates and scientific terminology, distinguishing between weak and strong AI, as well as functional and phenomenal consciousness, to offer a useful classification of AC systems.

Second, we explore how current AC systems work and what they are capable of. This section highlights the state of the art in AC research, focusing more specifically on the recent synergy between two scientific models of consciousness – the Global Workspace and Attention Schema. We also discuss the problem of testing for consciousness. The section concludes that while current technical implementations are most likely not conscious in the human sense, some metaphysical views leave room for the possibility that artificial agents could already possess a form of subjective experience.

Having laid these foundations, the third section addresses the ethical challenges posed by AC. We seek to provide a balanced overview of the risks and opportunities, reviewing existing literature and without advocating for a specific stance. A central ethical issue is how AC systems should be treated, particularly in relation to moral status. Furthermore, AC is expected to amplify societal many of the societal impacts already associated with AI, both positive and negative. We conclude optimistically by noting that while the development of AC appears inevitable, its trajectory can be steered through deliberate and responsible efforts. Importantly, this path is essential to unlock significant scientific discoveries.

Nevertheless, balancing these risks and opportunities is a challenging task. We propose six main recommendations for the responsible development of AC. Lastly, we highlight some limitations of the current study. Notably, AC research will increasingly merge with other new technologies, besides current AI. The foreseeable magnitude of the impact of AC pressingly requires more efforts to guide its development responsibly and transparently.

# 2 Artificial Consciousness: An Etymology

Research in artificial consciousness is a young but highly interdisciplinary field. As such, it inherits a rich epistemic domain shaped by multiple disciplines (Elamrani & Yampolskiy, 2019). However, its terminology remains ambiguous and heterogeneous, often leading to misunderstandings even among experts. This section builds key definitions and concepts etymologically, by tracing their historical development.

In Section 2.1, we examine how the rise of AI coincided with a computational turn in our understanding of the human mind. This shift underscored the importance of consciousness and led to the birth of *consciousness science*, where artificial systems now play a significant role.

In Section **Error! Reference source not found.**, we define key terms based on their historical context and highlight pressing scientific issues through an analysis of key distinctions.

## 2.1 Scientific History

### 2.1.1 Artificial Intelligence

The history of artificial intelligence is entwined with the birth of information theory, computation theory, neuroscience and cybernetics, fields that together sought to decode the languages of mind and machine. These disciplines converged on a shared ambition: to understand and replicate the principles that govern intelligent behaviour.

The birth of neural networks owes much to the union of information theory with the study of brain activity. As scientists trained their attention on the electrical signals pulsing through neural pathways, tools such as EEGs (Berger, 1929) became the translators of a hidden language, the oscillations of thought itself. It was here, amid efforts to process and model these signals, that the first artificial neurons took root—rudimentary, but bold in their imitation of biological counterparts (Mcculloch & Pitts, 1943).

The emergence of the electronic computer, underpinned by computability theory (Turing, 1936), transformed this dream into a practical endeavour. Machines capable of executing algorithms became, for the first time, instruments of both simulation and design. The moment was crystallised at the Dartmouth Summer Research Project on Artificial Intelligence in 1956 (McCarthy et al., 1955), where the term "Artificial Intelligence" was coined, and the idea of teaching machines to think entered the cultural lexicon.

Herbert Simon, in "The Sciences of the Artificial" (Simon, 1969), argued that artificial systems deserved a scientific lens of their own. Intelligence, he proposed, was not merely a natural phenomenon but an engineered process, one capable of reshaping the boundaries of problem-solving. His vision was an invitation to see the artificial not as an imitation of the natural, but as its creative counterpart. He set the boundaries for sciences of the artificial through the following points:

1. *"Artificial things are synthesized (though not always or usually with full forethought) by human beings.*
2. *Artificial things may imitate appearances in natural things while lacking, in on or many respects, the reality of the latter.*
3. *Artificial things can be characterised in terms of functions, goals, adaptation.*
4. *Artificial things are often discussed, particularly when they are being designed, in terms of imperatives as well as descriptives."*

Meanwhile, cybernetics, the study of control and communication in information processing systems, offered a complementary framework. With feedback at its core, cybernetics framed agency as a matter of self-regulation—information looping through sensors, processors, and actuators, whether in biological organisms or machines (Wiener, 1965). It was an elegant theory of systems that adapt and learn, rooted in the flow and balance of information.

As these ideas grew, so did the opposition between two distinct visions of machine intelligence. *Strong AI* seeks to *realise* mental states within electronic computer systems. This term refers to systems that genuinely possess understanding and other cognitive states. In this view, "*given the*

*right programs computers can be said to have understanding and other cognitive states*" (Searle, 1980). In contrast, *Weak AI* merely seeks to *simulate* minds in computer systems, providing a useful tool for the analysis of mind, but without any pretension to ever achieve the sort of genuine mental states present in humans. Under this view, "*the principal value of the computer in the study of the mind is that it gives us a very powerful tool. For example, it enables us to formulate and test hypotheses in a more rigorous and precise fashion.*" (Searle, 1980).

This influential distinction was famously illustrated by the "Chinese Room" thought experiment. In this scenario, a person inside the room manipulates Chinese characters based on a set of instructions, piecing together sequences that appear meaningful to an outside observer who masters the language. However, the person inside the room *themselves* have no knowledge of Chinese: they cannot discriminate the symbols, nor meaningfully relate them to referents in the real world. Although they manage to produce a meaningful answer to the question asked in a language foreign to them, they never understood a word of the conversation. Searle argues that a computer program, much like the person in the room, can only manipulate symbols by following instructions, but that no "*understanding*" ever happens in software.

The symbolic representation available in the Chinese Room is *simulacral*: it operates in a simulation completely detached from reality (Chalmers, 2022, pp. 23–24). Similarly, Weak AI operates within its own simulacral domain. It processes patterns, symbols, or data in a way that simulates intelligent behaviour, but it does so without engaging with our representation of reality. In contrast, Strong AI is *performative*: it engages with the world on the same representational level as humans, it operates over the same domain.

A similar distinction holds in the case of AC, as we will discuss in more details in section 2.2.3 Strong vs Weak, Phenomenal vs Functional: Classifying AC.

### 2.1.2 The Science of Consciousness

The scientific study of consciousness lies at the crossroads of consciousness science, philosophy of mind, and cognitive science, drawing insights from disciplines as diverse as neuroscience, psychology, and computer science. Its ambition is as formidable as its subject: to understand the nature of conscious experience and its relation to the physical and functional processes of the mind.

Consciousness resists a singular definition, its complexity evident in the variety of approaches that attempt to capture it. Informally, it is often described as the phenomenon that begins upon waking up and ceases during dreamless sleep or coma. More formally, Ned Block delineated a popular distinction between *phenomenal consciousness*, which refers to the subjective quality of experience —what it feels like to see red or hear a melody— and *access consciousness*, narrowly defined as the processes by which information becomes available for reasoning, reporting, and decision-making (Block, 1995).

While Block's notion of access consciousness has evolved over time and taken on additional connotations (Naccache, 2018; Schlicht, 2012), the more compelling distinction it highlights is between the subjective, qualitative nature of phenomenal consciousness and the objective, functional aspects of consciousness – the computation underlying the phenomenon. This distinction provides a foundation for understanding two essential dimensions of consciousness:

- **Phenomenal consciousness:** the subjective quality of experience, or "what it feels like".
- **Functional consciousness:** the computational mechanisms underlying conscious information processing.

This fundamental dichotomy between functional and phenomenal aspects of consciousness came to prominence through early computational theories of mind, ultimately crystallising in what Chalmers famously coined the *hard problem of consciousness* (Chalmers, 1996), (discussed below in section 2.1.2.1). This conceptual distinction coincided with the start of a rigorous scientific investigation of consciousness (below in section 2.1.2.2).

### 2.1.2.1  *Functionalism and Computational Theories of Mind*

The cognitive turn of the mid-20th century sparked a profound shift in the study of mind. Computation theory, through groundbreaking concepts such as Turing's universal computing machine (Turing, 1936), established a new rigorous analytical framework for understanding mental states in terms of information processing. The computational paradigm in the study of mind quickly permeated disciplines as diverse as neuroscience, psychology, economics, social sciences, and linguistics.

In philosophy, this epistemological shift gave rise to functionalism, a theory which classically argues that mental states, such as pain, are best defined by their functional role than their physical instantiation (Putnam, 1967). Today, such computational theories of mind, built around the analogy of the mind as software and the brain as hardware, are the most popular analytical approach in the study of our mental lives (Rescorla, 2020). One key implication of these views is *multiple realisability*: the idea that a same mental state may be realised by different physical architectures. Theoretically, this opens the possibility for electronic computers to experience mental states similar to ours. More importantly, it is challenging fundamental assumptions about the very nature of scientific knowledge (Fodor, 1974; Kim, 1992).

While computational theories of mind brought remarkable progress in understanding cognition, they also illuminated a persistent problem: the *explanatory gap* between physical or functional descriptions of the mind and the qualitative, subjective nature of experience (Levine, 1983). What does it feel like to perceive a vivid red, resonate with a melody, or experience pain? This "*what it is like*" dimension of consciousness, its *phenomenal quality*, resists reductive physical explanations (Chalmers, 1996; Jackson, 1982; Nagel, 1974).

Through a series of thought experiments, Ned Block's powerfully illustrated the limitations of functionalism (Block, 1978). Consider, for instance, a homunculi-headed robot, a being whose brain consists of tiny agents (homunculi) performing computational tasks that replicate the functional roles of neurons. Or, alternatively, imagine a configuration where a hundred billion of people are assigned the role of a neuron, and through binary radio signals, they collectively copy the functional organisation of a human brain. Block compellingly argues that although such systems may successfully replicate neuronal functions with extraordinary precision, it remains implausible and misguided to claim that they are *phenomenally* conscious. Further thought experiments, like zombies (Chalmers, 1996; Kirk, 1974), brought this inadequacy into sharp relief. A *philosophical zombie*, functionally isomorphic to a human, yet devoid of inner experience, challenges the functional role of phenomenal consciousness if information processing and relevant behaviour can occur without it. These puzzles highlight a gap between conscious information processing and subjective experience.

At its heart, the issue of phenomenal consciousness is best captured by Chalmers' formulation of the *hard problem* of consciousness (Chalmers, 1996): why is conscious information processing not all happening "*in the dark*"? Despite significant progress in understanding the brain as an information-processing system, the leap from computation to qualitative experience remains one of the most puzzling issue of science and philosophy.

### 2.1.2.2 The Birth of Consciousness Science

For centuries, consciousness remained an elusive, philosophical or speculative domain of study. But the late 20th century witnessed the rise of a concerted scientific effort to investigate its place in nature. This shift was driven by a series of groundbreaking theories that began to delineate consciousness as an empirically accessible phenomenon.

Among the earliest and most influential of these was the Global Workspace Theory (GWT) (Baars, 1988). According to GWT, consciousness functions as the spotlight of mental activity, highlighting the most salient information from a multitude of neuronal subsystems to make it available globally for reasoning and decision-making. Around the same time, David Rosenthal developed the Higher-Order Thought (HOT) theory, linking consciousness to the brain's reflective ability to represent its own states (D. M. Rosenthal, 1986). Meanwhile, Gerald Edelman developed an approach mixing insights from both computation theory and biology, emphasising the role neuronal mappings, evolutionary constraints, and dynamic interaction between the brain and body in shaping conscious experience (Edelman, 1990).

Philosophers were equally engaged in reframing the problem of consciousness and offered contrasting frameworks for understanding consciousness. Daniel Dennett argued for a fully materialist account, introducing the intentional stance, which interprets systems as having beliefs and desires based on their functional organisation (Dennett, 1988). His multiple drafts model rejected the idea of a central "cartesian theatre" of consciousness, describing it instead as a decentralised process, in which multiple competing narratives are continuously built and updated across the brain (Dennett, 1991). In contrast, David Chalmers (Chalmers, 1996) defended naturalistic dualism, which asserts that while consciousness is part of the natural order, it cannot be reduced to physical processes alone. He argued that *qualia*[1] must be explained through fundamental principles that escape traditional materialist frameworks.

Thus, by the 1990s, theoretical advances were converging to form the foundation of a robust science of consciousness. Researchers started discussing empirically testable models or conditions, setting the stage for a new era of interdisciplinary inquiry. At that time, consciousness science established itself as a distinct and thriving field. Dedicated conferences, such as the Science of Consciousness[2] conference, and organisations like the Association for the Scientific Study of Consciousness[3], provided platforms for interdisciplinary collaboration. New peer-reviewed journals such as the Journal of Consciousness Studies[4] brought together experimental, theoretical, and philosophical work, consolidating a strong scientific community.

---

[1] Singular: quale; refers to the qualitative properties of phenomenal experience.

The start of the 21st century was marked by a surge of experimental discoveries across neuroscience, psychology, and related disciplines, providing unprecedented insights into the mechanisms underlying consciousness. These studies significantly advanced our understanding of the so-called neural correlates of consciousness (NCCs) – specific patterns of brain activity associated with conscious experience. Using computational tools like fMRI, EEG, and neural decoding algorithms, researchers began mapping scientific predictions with the dynamic interplay of neural networks that support phenomena such as perception, attention, and self-awareness. This confluence between empirical data and computational models grounded new scientific accounts of consciousness.

Among them, Integrated Information Theory (IIT) (Tononi, 2004), states that consciousness correlates with the ability for a system to integrate information. This is captured through the metric $\Phi$, which provides a quantified, yet non computable, measure of the level of information integration within a system, based on complexity theory.

Likewise, Predictive Processing (PP) is rooted in computational models of brain function (Clark, 2013; K. Friston, 2010; Hohwy, 2014). This framework conceptualise the brain as a prediction engine, constantly generating probabilistic models to anticipate sensory inputs and normalise prediction errors (K. Friston, 2005; A. K. Seth, 2015). Active inference extends this by integrating action, showing how organisms actively reshape their environments to reduce uncertainty (K. J. Friston et al., 2014; Pezzulo et al., 2015).

Meanwhile, the Attention Schema Theory (AST) provides another influential computational approach of consciousness, mixing evolutionary insights from social cognition with bodily illusions such as evidenced through the famous rubber hand experiment (Graziano, 2013). The resulting view claims that phenomenal consciousness results from the brain's ability to create an *attention schema*: a simplified, monitoring model of its own attentional processes. Computational simulations of attention dynamics have been instrumental in testing and refining this theory supporting a functionalist view of awareness (Graziano, 2017, 2022).

---

[2] https://consciousness.arizona.edu/science-consciousness-conferences
[3] https://theassc.org/about-us/#history
[4] https://en.wikipedia.org/wiki/Journal_of_Consciousness_Studies

These and many other approaches, as catalogued by Seth and Bayne (A. K. Seth & Bayne, 2022), highlight the indispensable role of computational tools and models in consciousness science. By simulating neural systems, quantifying informational structures, and normalising predictive mechanisms, computational approaches provide a framework for connecting the physical processes of the brain with the rich, subjective experience of the mind. Together, these theories represent an interdisciplinary effort to build a scientific account of consciousness, blending experimental data with computation modelling.

### 2.1.3 Early History of Artificial Consciousness

By the late 90s, researchers started tackling the question of consciousness in artificial systems more openly. This trend was more markedly established through a workshop entitled "Can a Machine be Conscious", held in May 2001 at the Banbury Center, Cold Spring Harbor Laboratory, bringing together leading figures of the field[5]. The general consensus resulting from this meeting was that **_"in principle, one day computers or robots could be conscious"_** (Koch, 2001).

The field was always interdisciplinary, and so were the first approaches to build consciousness in artificial agents. While some implementations focused on robotic applications (Alami et al., 2006; Chella & Macaluso, 2006; Holland, 2003; Macaluso et al., 2007), others were instead more focused on cognitive architectures (Aleksander, 2005; Haikonen, 2003; Sloman & Chrisley, 2003). Ultimately, a dedicated journal was created in 2009, the "International Journal of Machine Consciousness" (Chella & Manzotti, 2009), recently rebranded as the "Journal of Artificial Intelligence and Consciousness" (Chella, 2020).

Two main motivations have always driven the field. The first one is that artificial consciousness is a remarkable tool to decipher natural consciousness, with the potential to ultimately unveil the mystery of phenomenal experience. The second one is that, artificial consciousness paves the way for the ultimate kind of artificial intelligence, one modelled closely after our own cognition, and therefore capable of the most relevant interactions with the natural world.

---

[5] http://www.theswartzfoundation.org/banbury_e.asp

## 2.2 Terminology

### 2.2.1 Artificial Consciousness

As discussed above, artificial consciousness is born from the intersection of artificial intelligence and the scientific study of consciousness. Following Herbert Simon's definition of "artificial" as systems designed to mimic or extend natural phenomena (see above, 2.1.1), artificial consciousness can be broadly defined as either:

**A technology:** AI systems that imitate the appearance of conscious information processing, without necessarily achieving the reality of the natural phenomenon.

**A scientific field:** The interdisciplinary study of such systems, their mechanisms, and their implications, both practical and philosophical.

At its core, artificial consciousness attempts to address the "inner life" of a system—what it might mean for an AI to possess subjective experience or to function as though it does. This makes of AC not only a technical endeavour but also, and more importantly, a great scientific and ethical challenge.

AC cannot be seen as a monolithic effort following a smooth and consensual development path. Rather, progress in AC tends to profuse chaotically. Indeed, consciousness involves several mental features and cognitive abilities, and this plurality is reflected in the variety of approaches and frameworks that explicitly address AC. For instance, a meta-analysis of AC tests alphabetically listed the following central features from reviewed literature: *attention, awareness, creativity, dynamism, emotions, grounding, imagination, intelligence, intentionality, language, qualia, perception, self,* and *volition* (Elamrani & Yampolskiy, 2019). The same study found that such approaches were typically influenced by the scientific background of the researchers (e.g. engineers would focus more on "self" while neuroscientists would focus more on "qualia"). We should note that to the exception of qualia, most of these features have already been deliberately implemented to some extent in AI systems. And in fact, progress in AC is not confined to explicit efforts in this direction, but may also be driven by AI development.

### 2.2.2 AI vs AC

The relationship between consciousness and intelligence runs deep. Correspondingly, a faithful classification of systems along this spectrum would be more gradual than strictly binary. Although there is no clear demarcation between AI and AC, some AI systems align more closely with the AC programme than others.

Indeed, within AI, one can distinguish between two main historical branches. The first one is more logic and heuristic based[6], being predominantly task-oriented and driven by specific problem solving, such as banking systems, logistics or verification. These systems, while sophisticated, are narrowly focused and rely on ad-hoc algorithms tailored to external objectives. The other branch, by contrast, draws inspiration from natural minds, incorporating neuro-inspired features like attention, memory, and learning, alongside efforts to develop multi-modal systems capable of integrating diverse forms of input (e.g., text, image, and audio). These latter systems, with their broad and adaptive capabilities, grounded in cognitive and neuroscientific insights, align more closely with the domain of artificial consciousness.

This underscores that artificial consciousness is not confined to explicit computational implementations of scientific models of consciousness in AI. Instead, it encapsulates a broader shift towards designing systems that imitate natural cognitive architectures. Nevertheless, a strong driver of AC consists of explicit implementations of computational models of consciousness in AI systems.

### 2.2.3 Strong vs Weak, Phenomenal vs Functional: Classifying AC

Artificial consciousness inherits two key distinctions from its parent disciplines: *Strong* vs *Weak* AI (see above section 2.1.1) and *Phenomenal* vs *Functional* consciousness (see above section 2.1.2.1). By combining these distinctions, we arrive at a further classification for AC (Holland, 2003; A. Seth, 2009):

**Strong AC:** AI with phenomenal consciousness.

**Weak AC:** AI with functional or simulated consciousness, but without experience (i.e., they are, at best, "zombies" in philosophical terms – see section 2.1.2.1 above).

---

[6] It is sometimes referred to as "GOFAI" for "Good Old Fashioned AI", since it was the predominant approach during the first AI boom, when computing power was scarce.

For more clarity, one can further cross the Strong/Weak distinction with the Functional/Phenomenal one, highlighting four categories for classifying AC systems as shown below (Table 1).

|  | **Functional Consciousness** | **Phenomenal Consciousness** |
|---|---|---|
| **Strong AI (Performative)** | **Strong Functional AC** If full functional performance entails phenomenal experience, these are *Strong AC* systems; if not, they are *Weak AC*. | **Strong Phenomenal AC** *Strong AC*: AI systems with genuine inner life. |
| **Weak AI (Simulacral)** | **Weak Functional AC** *Weak AC*: Systems that simulate functional access without instantiating it. | **Weak Phenomenal AC** *Weak AC:* Simulations of qualitative states without experience. |

*Table 1: Classifying AC Systems*

1. **Weak Functional AC:** Systems computing functions that are typically associated with consciousness (e.g. attention, creativity, or imagination). While capable of simulating advanced cognitive features and behaviours, such systems have no phenomenal experience: they are instances of *Weak AC*.
2. **Weak Phenomenal AC:** Systems falling in this class simulate phenomenal experience, although they do not genuinely experience conscious states. For instance, these could be computational models aiming to map out qualia space[7] or used to make testable predictions about phenomenal experiences without instantiating them. But there is nothing it feels like to be such a system: although it simulates a phenomenal space, it remains a *Weak AC*.
3. **Strong Functional AC:** These systems compute the functions underlying conscious information processing, operating on the same functional level as humans. This class underscores a critical issue: is full functional replication sufficient to spark phenomenal consciousness or could such systems remain philosophical zombies? If performing the right functions is sufficient for phenomenal experience, these systems automatically fall under the category of *Strong AC*. However, the functional aspect could be necessary but not sufficient. An additional condition, such as being implemented in a biological substrate, might prevent such systems from reaching phenomenal states. In which case, this class of systems belongs to *Weak AC*.

---

[7] The idea of mapping qualia, or what-it-feels like experience, to a mathematical space has been notably developed by Aaron Sloman (Sloman, 1993) and Giulio Tononi (Tononi, 2004).

4. **Strong Phenomenal AC:** These are Strong AI systems with phenomenal experience, and as such they are *Strong AC*. If computing a specific function is necessary and sufficient for phenomenal experience, independently of any other consideration such as physical substrate, then this class is equivalent to the previous one. However, if phenomenal experience is dependent upon a non-functional element, Strong Phenomenal AC is a separate class.

While most of these classes accept a straightforward classification between Strong and Weak AC as defined above, one of them – Strong AI / Functional Consciousness – hosts an ambiguity which highlights the deepest boundaries of our scientific knowledge. Does phenomenal experience ultimately boils down to functions? Could a natural phenomenon irremediably escape a computational characterisation?

Whether and to what extent Strong AC is achievable is strongly tied to the metaphysical configuration of the world: what are the constitutive elements of reality and how do they interact to form what we know?

Among the most permissive views are panexperientialist theories (Chalmers, 2016a). According to them, phenomenal experience is ubiquitous (e.g. found even in plants, thermostat or atoms). suggesting that qualia are as fundamental as mass or electric charge. Applied to AI, they suggest that current AI systems are at least proto-conscious[8]. One key challenge such views raise is the *combination problem* (Chalmers, 2016b). In the case of AC, it amounts to answering whether, and how, AI systems can combine these proto-conscious states into a unified phenomenal experience.

In contrast, the most restrictive views tend to be anthropocentric and will suggest some form of biological substrate-dependence for the realisation of phenomenal states, excluding the possibility of Strong Phenomenal AC in electronic computers (Godfrey-Smith, 2016; A. Seth, 2024).

---

[8] Chalmers defines proto-conscious properties as "special properties that are precursors to consciousness and that can collectively constitute consciousness in larger systems" (Chalmers, 2016a)

### 2.2.4 Artificial Consciousness vs Machine Consciousness

Since the inception of AI, the term "machine intelligence" (Turing, 1950) has been synonymous with "artificial intelligence", and similarly, "machine consciousness" is often used interchangeably with "artificial consciousness". While this overlap is generally unproblematic, it can be helpful to distinguish between the two in certain contexts. The term "machine consciousness" may be preferred for the mechanistic or physical aspects of consciousness, focusing on its potential embodiment in hardware or engineered systems. In contrast, "artificial consciousness" maintains a clear conceptual connection to AI (A. Seth, 2009), emphasising the disembodied, algorithmic, software aspect of conscious information processing systems.

### 2.2.5 Strong AC vs Synthetic Phenomenology

One useful terminological clarification could be drawn between Strong AC and Synthetic Phenomenology. Although both terms refer to engineered phenomenal properties, it could be convenient to not use them as completely synonymous, and reserve each of them for a slightly specific meaning. As highlighted above, Strong AC maintains an explicit continuity with classic AI terminology (A. Seth, 2009). In contrast, Synthetic Phenomenology is not related to a specific technology and can therefore encapsulate any substrate: it is not restricted to software running on electronic devices, but also includes brain organoids, quantum computing, etc. We can foresee that in a future where the science of consciousness will intersect with new technologies, further terminological distinctions will ensue. By adopting this distinction, we rid the field of a redundancy and create more accurate definitions:

- **Synthetic Phenomenology:** umbrella term for the engineering of phenomenal properties within any substrate, through any form of technology.
- **Strong Artificial Consciousness:** synthetic phenomenology within AI systems, maintaining a continuity with classic terminology.

## 3 State of The Art of Artificial Consciousness

As discussed in the previous section, the beginning of artificial consciousness (AC) as a distinctive field of study is a natural progression from the respective maturity of artificial intelligence (AI) and consciousness science. Both parents disciplines share a foundation in computational models, and their ongoing convergence has given rise to new possibilities for imitating conscious information processes in artificial systems. In this section, we focus on more recent trends in AC, and notably those that follow the recent AI revolution brought by deep learning architectures in the past decade.

## 3.1 The Catalytic Convergence of AI and Neuroscience

From its earliest days, AI has drawn inspiration from neuroscience, implementing computational models of the mind in digital systems (see 2.1.1 above). Neural networks, for instance, are loosely based on the architecture of biological brains. Conversely, AI has continuously informed neuroscience, offering tools and frameworks for understanding the computational underpinnings of perception, learning, and memory. Recently, an important development came with the rise of Deep Learning (DL), which demonstrated the ability to automatically learn features—regular patterns of information—without explicit programming (Hinton, 2014; LeCun et al., 2015; Schmidhuber, 2015). DL systems revolutionised tasks like image classification, achieving results comparable to human performance. This breakthrough revealed striking similarities between artificial and natural systems and further integrated research between AI and neuroscience (Caucheteux et al., 2022; Hassabis et al., 2017; Savage, 2019; J. Wang et al., 2022).

The increasing convergence of these fields is exemplified by interdisciplinary forums like NeurIPS (conference on NEURal Information Processing Systems), which fosters collaboration at the intersection of AI, neuroscience, and cognitive science. Modern architectures such as transformers and attention mechanisms have further deepened this relationship (Vaswani et al., 2017), enabling more effective modelling of dynamic, context-sensitive information—a capability that aligns with both biological cognition and theories of consciousness.

## 3.2 Computational Models of Consciousness

In parallel with these advances, neuroscience started integrating consciousness research. Historically relegated to abstract metaphysical debates, the scientific study of consciousness gained credibility through the development of formal models and experimental tools. Computational approaches have played a critical role in this transformation (see 2.1.2 above).

A cornerstone of this research is the distinction between conscious and unconscious processing in the brain. The human brain processes vast amounts of information, but only a fraction reaches conscious awareness. Over the years, accumulating evidence from experimental studies in neuroscience, cognitive science and psychology have shown that this selection is context-dependent, influenced by factors like saliency (e.g., detecting edges), criticality (e.g., detecting threats), and availability (e.g., recent memory). Despite several competing theories (A. K. Seth & Bayne, 2022), most scientific accounts of consciousness characterise it as an information process, often emphasising its role in enhancing adaptive behaviour. Consciousness is often, although not universally, seen as an evolutionary adaptation, granting organisms an advantage in decision-making, perception, and memory. These function-oriented approaches are easily transposable to computer programs.

### 3.3 The Consciousness Prior

More recently, the convergence of AI and consciousness science was significantly shaped by the notion of a *consciousness prior* by Yoshua Bengio (Bengio, 2019). This concept proposes an innovative approach to representation learning inspired by cognitive psychology and neuroscience.

The first source of inspiration for the *consciousness prior* is a theory developed by Amos Tversky and Daniel Kahneman, and popularised in the book "*Thinking, Fast and Slow*" (Kahneman, 2012), which divides mental operations into two systems:

- *System 1* is fast, automatic, and largely unconscious. It processes information effortlessly, relying on heuristics for quick responses, and governs everyday activities like recognising faces, interpreting emotions, or navigating familiar environments.
- *System 2*, in contrast, is slow, deliberate, and effortful. It requires conscious attention and is engaged in tasks demanding focus, such as solving mathematical equations or evaluating conflicting evidence. According to Bengio, "*Typical System 2 tasks require a sequence of conscious steps, which also means that they tend to take more time than System 1 tasks. By this definition, System 2 abilities are closely related to consciousness.*" (Bengio, 2019)

The second source of inspiration for the *prior* comes from one of the leading scientific accounts of consciousness: the Global Workspace Theory (GWT) (Baars, 1988; Dehaene et al., 1998; Dehaene & Changeux, 2011; Dehaene & Naccache, 2001), which conceptualises consciousness as a central workspace where selected information becomes globally accessible across multiple specialised cognitive subsystems. This framework provided the *consciousness* prior with a model for how conscious processes could be computationally implemented.

Building on these insights, Bengio characterises consciousness functionally, as a bottleneck for information processing that prioritises relevant inputs while filtering out distractions. Under his view, conscious processing enables the discovery of high-level abstractions, improving representations and facilitating collaborative task-solving. He furthermore suggests that consciousness allows agents to communicate their high-level states more effectively through language, making them more interpretable.

The introduction of the *consciousness prior* represents an important step in the advancement of AC. Not only does it reinforce the idea that modelling AI more closely after human cognition can yield better results, but also, it invigorates research by explicitly embedding insights from consciousness science into AI development.

## 3.4 Current AC Implementations

The most straightforward implementations of AC involve integrating scientific models of consciousness within AI systems. In this section, we offer a concise overview of the state of the art, deliberately avoiding overly technical details. The first part highlights one of the most actively developed approaches, intersecting Global Workspace Theory (GWT) and Attention Schema Theory (AST). The second part briefly reviews additional examples to illustrate current the diversity of perspectives and methodologies in the field. The aim is to convey that AC is not only a rich and multifaceted domain of research but also one with significant open questions and untapped potential for further exploration.

### 3.4.1 The GWT-AST Synergy

A notable recent trend in AC is the refinement of AI models by integrating insights from the Global Workspace Theory (GWT) and the Attention Schema Theory (AST), two complementary frameworks for understanding the function of consciousness.

The origins of GWT can be traced back to the work of Bernard Baars in 1988, who proposed that consciousness is a broadcast of information across a global workspace of cognitive systems (Baars,

1988). This theory was further developed into a neuronal model by Stanislas Dehaene, providing a detailed neural implementation of GWT and framing it as a central framework for understanding conscious processing (Dehaene et al., 1998). Over the next decade, GWT was refined through both experimental and theoretical approaches, building a computational account of conscious processing (Dehaene, 2014; Dehaene et al., 2014; Dehaene & Changeux, 2011; Dehaene & Naccache, 2001).

Meanwhile, motor approaches to consciousness were being developed, taking action control as a starting point. Among them, the passive frame theory and the attention schema theory were shown compatible and complementary (Morsella, 2005; Graziano, 2013; Morsella et al., 2016; Graziano & Morsella, 2019). According to the AST, consciousness evolved as a mechanism for monitoring and controlling attention. Drawing inspiration from the concept of a *body schema*, the AST postulates that the brain constructs an internal model of its attention processes—an "attention schema"—allowing for the efficient allocation of cognitive resources. The AST's mechanistic account of subjective experience is argued to provide a foundation for engineering consciousness. A key strength of the AST lies in its testable predictions, such as the hypothesis that monitoring attention improves performance (Graziano & Webb, 2015; Graziano, 2017).

Although they differ, it would be wrong to understand AST and GWT as two competing theories of consciousness. In fact, their complementarity was discussed in a special issue of Cognitive Neuropsychology where a standard model of consciousness was presented (Graziano et al., 2020). According to this proposal, AST handles the selection and monitoring of attention, while GWT describes how selected information is globally broadcast for integration by cognitive systems. This synthesis was widely supported by leading scholars through invited commentaries, consolidating the initiative (Dennett, 2020; Frankish, 2020; Blackmore, 2020; Panagiotaropoulos et al., 2020).

In AI, the introduction of the c*onsciousness prior,* building on GWT, gave a new impulse to artificial consciousness (Bengio, 2019). Although not directly tied to AST, this prior highlights how attention mechanisms can act as bottlenecks for high-level cognitive processing, emphasising the importance of consciousness-inspired architectures in AC development. GWT-inspired models have also been argued to enhance modular AI systems, particularly for multimodal tasks, through the implementation of a Global Latent Workspace (VanRullen & Kanai, 2021).

Several practical implementations followed. For instance, the first experiment involving an attention schema within a neural network agent was designed to control visuospatial attention, significantly improving learning performance (Wilterson & Graziano, 2021). GWT has been applied to enhance coordination within modular AI systems (Goyal et al., 2022). Recent work has demonstrated how incorporating internal attention control in multi-agent reinforcement learning improves coordination across agents toward common goals (Liu et al., 2023). These developments suggest promising applications, such as enabling agents to monitor their own attention and that of others, leading to improved communication, value alignment, and deep focus (Liu et al., 2023). Carrying this approach forward, the EIC-funded project ASTOUND seeks to enhance the social abilities of chatbots by implementing an AST module.

While none of these implementations claim to perform the kind of conscious information processing found in humans, they represent valuable steps forward, pointing to multiple potential development paths. It is important to note that these implementations do not aim to replicate human cognitive functions in their entirety. Moreover, the concept of attention itself —and which is central to the AST— differs across the domains of psychology, neuroscience, and machine learning, adding further complexity to the task (Lindsay, 2020).

Despite these challenges, significant opportunities remain to unlock the full potential of AST and GWT in artificial agents. Much work lies ahead, but the progress made so far underscores the rich and fruitful possibilities in integrating these theories into AC development.

*Table 2- A Timeline of the GWT-AST synergy in AC*

| 1988-2014 | **Foundations of GWT** |
|---|---|
| | • Global Workspace Theory (GWT) was introduced by Bernard Baars (1988) and later developed into a neuronal model by Dehaene et al. (1998). GWT posits that consciousness arises from the global broadcasting of selected information to multiple cognitive systems, enabling integration and coordination.<br>• Further elaborations on GWT, including Dehaene's "Le Code de la Conscience" (2014) and various computational models (Dehaene et al., 2014), established GWT as one of the dominant theoretical frameworks for explaining conscious processing. |
| 2013-2017 | **AST and its Role in AC** |
| | • Attention Schema Theory (AST) was proposed by Michael Graziano in 2013. AST suggests that consciousness evolved as a mechanism to monitor and control attention, much like the body schema helps in controlling movement.<br>• Graziano and Webb (2015) further developed AST as a mechanistic account of subjective awareness, arguing that attention without awareness is possible but suffers from deficits in control, making the theory both testable and applicable to AC. |
|  | • By 2017, Graziano suggested that AST and GWT were compatible, but GWT alone was incomplete without the attention schema, which offers an explanation for subjective experience. |
| 2019 | **The Consciousness Prior** |
|  | Yoshua Bengio introduced the "Consciousness Prior", proposing that consciousness serves as a bottleneck for high-level cognitive processing. Bengio's framework builds on GWT by describing how attention mechanisms select information from a broader pool before broadcasting it for further decision-making and perception. While distinct from AST, this work shares the emphasis on attention control and information filtering. |
| 2020 | **AST-GWT Reconciliation** |
|  | • Graziano et al. (2020) published a key paper reconciling AST, GWT, and other theories like higher-order thought and illusionism. This work demonstrated that AST and GWT are highly compatible, with AST handling attention selection and GWT handling the global broadcasting of that selected information.<br>• This reconciliation was widely supported by influential figures such as Dennett, Frankish, and Dehaene's team, marking a significant step toward a unified model of consciousness applicable to both biological and artificial systems. |
| 2021 | **First Computational Implementation of AST** |
|  | • Wilterson & Graziano (2021) developed the first computational implementation of AST using a deep Q-learning neural network agent. This agent, designed to perform a visuospatial task, demonstrated that an attention schema significantly improves learning and performance, supporting the idea that AST plays a crucial role in attention control.<br>• Experiments showed that without the attention schema, the agent's performance in attention control dropped significantly, reinforcing the practical benefit of incorporating AST into AI systems. |
| 2021-2022 | **GWT in Deep Learning and Modular AI Systems** |
|  | • In 2021, VanRullen & Kanai explored the application of GWT in deep learning, showing that neuro-cognitive architectures like GWT could improve machine learning by providing robustness and flexibility in modular AI systems.<br>• Bengio's team (2022) expanded this approach by proposing a shared global workspace model for coordination among neural modules. This workspace model improved visual reasoning and task performance in AI systems by integrating GWT principles into neural architectures. |
| 2023 | **Application of AST in Neural Agents** |
|  | • Liu et al. (2023) implemented AST in neural agents to control attention in complex tasks. Their experiments showed that agents equipped with an attention schema demonstrated improved task performance, faster learning, and better adaptability. |

### 3.4.2   Other Implementations of AC

Beyond the GWT-AST synergy, where the direct goal is the computational implementation of scientific models of consciousness, recent years have witnessed a growing diversity in the approaches and implementations of AI systems inspired by neuroscience and aiming to imitate essential features associated with conscious information processing. Notable trends include improving their generalisation abilities by forming meaningful representations (Greff et al., 2020; Schölkopf et al., 2021), pushing toward autonomous machine intelligence with energy-driven models (Dawid & LeCun, 2024; LeCun, 2022), as well as developping meta-cognitive abilities (Kanai et al., 2024; Lake & Baroni, 2023; Langdon et al., 2022; J. X. Wang, 2021).

As noted earlier (see section 2.2.2 above), not all endeavours in AC are deliberate attempts at implementing consciousness in AI. As AI and neuroscience progress, so does the risk of accidentally implementing AC. To address this concern, a recent study proposed several indicators of consciousness derived from leading theories to assess the presence of conscious-like processes in artificial systems (Butlin et al., 2023). These indicators were applied to a diverse set of case studies, including Transformer-based Large Language Models (Vaswani et al., 2017), the Perceiver architecture (Jaegle et al., 2021; Juliani et al., 2022), DeepMind's Adaptive Agent (Team et al., 2023), a virtual rodent agent (Merel et al., 2019), and PALM-e (Driess et al., 2023), an embodied multimodal language model. The selection of these case studies reflects the diversity of systems currently considered as potential candidates for AC, whether intentionally designed as such or not.

Another major source of potential for building AC systems comes from the increasing mathematisation of consciousness science, as evidenced by the growing success of the annual Mathematical Models of Consciousness conference, first held in 2019. This emerging community is dedicated to uncovering the mathematical structures underlying consciousness, exploring a diverse array of computational implementations. One prominent example is the Conscious Turing Machine, a framework that applies principles of theoretical computer science to define consciousness within the formal structure of a Turing machine (Blum & Blum, 2022). This approach emphasises computational concepts of space, time and energy to conscious-like behaviour. Another significant contribution is the Projective Consciousness Model, which takes a geometry-based approach, mapping conscious states onto structured mathematical spaces (Rudrauf, Sergeant-Perhtuis, et al., 2023; Rudrauf, Sergeant-Perthuis, et al., 2023). This model posits that phenomenal experiences correspond to specific geometric or topological features, suggesting a novel framework for understanding the relationship between structure and experience. Most of these models are driven by the structural turn in consciousness science (Fink et al., 2021; Kleiner, 2024), where it is assumed that phenomenal experience coherently maps to a structured mathematical space.

Overall, progress in AC highlights the computational nature of reality. Biological and artificial information processes alike decompose external physical signals mathematically, with their underlying structure robustly mapping to internal representations. This alignment holds promise for functional AC, but the ultimate question of strong AC remains unresolved.

## 3.5 The Problem of Testing

Tests for consciousness typically fall under one or both of the following categories: 1- behavioural approaches, such as adopted by Descartes[9], assessing information processing systems in terms of their actions (identifying conscious states by external inputs-outputs relations), 2- architectural approaches, such as put forward by Leibniz[10], assessing information processing systems in terms of their subjective representations (identifying conscious states by internal inputs-outputs relations)(Elamrani & Yampolskiy, 2019). For artificial consciousness, neither of these approaches are satisfactory on their own.

The typical behavioural test for human consciousness will involve verbal reports. Since humans share a similar physical architecture, it is reasonable to trust this method. But in the artificial case, behavioural tests derive from Turing's imitation game, where the algorithm needs to imitate human performance. As Turing said, the "*mysteries*" of consciousness "*need not be solved in order to answer the question we are concerned with*"(Turing, 1950). The obvious problem with this approach is that it might fail to identify *zombies*: AC systems that would behave like humans, yet would lack subjective experience.

In contrast, architecture-based approaches will involve examining the brain's activity through neuroscientific evidence, such as EEG or functional MRI scans, to correlate specific patterns of neural activity with conscious states. In the artificial case, such approaches attempt to evaluate whether the internal architecture of an AI system exhibits features thought to underlie consciousness, such as a global workspace, recursive self-modelling, or an attention schema. While architectural approaches avoid the pitfall of mistaking behavioural mimicry for genuine consciousness, they face significant challenges when applied to artificial systems. First, our understanding of the necessary and sufficient conditions for consciousness in any system—biological or artificial—is incomplete. Even if an AI's architecture imitates structures associated with human consciousness, this does not guarantee subjective experience. Second, architectural approaches risk anthropocentrism: they assume that features derived from human cognition (e.g., global workspaces) are universally necessary for consciousness.

---

[9] In "*Discours de la méthode*" (1637), Descartes argues that even if we could build machines capable of speaking, their behaviour would irremediably fail under some circumstances, thereby revealing their artificial nature. https://fr.wikisource.org/wiki/Page:%C5%92uvres_de_Descartes,_%C3%A9d._Cousin,_tome_I.djvu/192

[10] In "*La Monadologie*" (1714), paragraph §17, Leibniz argues that subjective experience is irreducible to matter by building the famous Mill thought experiment. His analysis focuses on the internal operations of a conscious machine – the pieces pushing against one another inside the mill, deliberately setting aside its external behaviour.

Thus, for artificial consciousness, neither behavioural nor architectural approaches alone are sufficient. Behavioural tests risk false positives by mistaking behaviour for consciousness, while architectural tests risk false negatives by ignoring possible non-human forms of conscious organisation. A robust test for artificial consciousness likely requires integrating both approaches while addressing their limitations.

One possibility is to develop hybrid methodologies that assess coherence between external behaviour and internal architecture. For example, a system could be tested for the presence of behavioural markers (such as adaptability, context-sensitive reasoning, or emotional simulation) while simultaneously verifying that they result from processes plausibly linked to subjective experience (Butlin et al., 2023; Pennartz et al., 2019). This might involve examining how the system models itself, its environment, and the interaction between the two.

Ultimately, developing reliable methods to detect or measure AC will require advances in philosophy, consciousness science, and AI research. It may also necessitate a rethinking of what it means to "test" for consciousness in the first place, as the concept might elude straightforward validation (Bayne et al., 2024).

### 3.6  Current Implementations of AC are Weak

The nascent state of AC is currently best characterised as Weak, with significant gaps remaining before the development of a truly Strong AC system. This is evident from the limitations of existing implementations, which, while fostering valuable interactions between AI and neuroscience, have yet to achieve the computational replication of any comprehensive theory of consciousness within an AI system.

Michael Graziano acknowledges that "*AST is not yet specific enough to hand a blueprint to an engineer*" (Graziano, 2017). Similarly, Yoshua Bengio clearly states that the consciousness prior is not directly addressing "*the notion of self and that of subjective perception*" but is rather aimed at "*the use of machine learning ideas and experiments as ways to formalize theories of consciousness*", "*identify advantages which they can bring to a learning agent*", "*and as a way to test these theories via machine learning experiments*" (Bengio, 2019). Rufin VanRullen and Ryota Kanai adopt a related position, where building a global workspace within deep learning architectures amounts to integrating into AI systems insights inspired from access-consciousness, as opposed to phenomenal consciousness (VanRullen & Kanai, 2021).

Moreover, Stanislas Dehaene, Hakwan Lau and Sid Kouider, emphasise that systems based on GWT would lack an additional self-monitoring aspect which is found in natural consciousness (Dehaene et al., 2017). And while such self-monitoring aspects could be, potentially, implemented via an AST, current computational implementations of an attention schema in AI are "*still far from being a rich, coherent, descriptive, and predictive model of attention*" (Liu et al., 2023).

The technological gap between engineered subjectivity and natural conscious processing is further evidenced by fundamental distinctions between attention in biological systems versus machine learning. In biological systems, attention operates through dynamic, hierarchical mechanisms, integrating sensory input with neuromodulatory influences such as dopamine and serotonin, which enable adaptive responses to environmental demands. These systems leverage complex nonlinear dynamics and plasticity across multiple scales, from synaptic adjustments to whole-network coordination, ensuring energy efficiency and robust adaptability. In contrast, artificial systems rely on static architectures, gradient-based learning, and computationally expensive processes, often requiring retraining to adapt to new contexts. (Cohen et al., 2022; Lindsay, 2020). Overall, these distinctions between natural and artificial neural networks highlight the limitations of current AI in replicating the flexible and integrative nature of biological cognition, underscoring the need for significant refinements of neuro-AI.

This divide reflects not only the technological challenges but also the early stage of the science of consciousness itself. Adversarial collaborations in consciousness studies have highlighted the lack of consensus on a comprehensive theory (Melloni et al., 2023), with significant experimental work still required to test leading frameworks like Higher Order Thought (D. Rosenthal, 2020) and meta-cognition (S. Fleming, 2023; S. M. Fleming & Lau, 2014). Such theories, concurrent with AST, will be critical for informing future attempts at computational modelling. Given this landscape, it would be exceedingly unlikely—if not impossible—to develop strong AC accidentally or without deliberate, theory-driven efforts (Chalmers, 2023).

However, the possibility of artificial agents with phenomenal experience cannot be ruled out so hastily. As mentioned in a previous section (2.2.3 Strong vs Weak, Phenomenal vs Functional: Classifying AC), under some extreme metaphysical views, everything in the universe has phenomenal experience—from atoms to planets, including electronic computers. Furthermore, functionalist approaches (see section 2.1.2 above) are still among the most popular contemporary theories of mind, despite a lot of criticism. According to such theories and others, artificial information processing systems could already have phenomenal experience.

The empirical road toward Strong AC remains a long and uncertain one, which will require a synthesis of theoretical insights, experimental validation, and technological innovation. While strides are being made, more interdisciplinary research will be necessary to fully tackle the issue. The result of this brief review of current AC systems shows that as far as the imitation of human consciousness is concerned, present implementations remain significantly incomplete.

# 4 Ethical Risks And Opportunities

Throughout history, major advances in science and technology have also reshaped our moral frameworks and societal structures. Since the advent of the computer revolution, the collective wellbeing of our societies is increasingly dependent on information and communication technologies (Floridi, 2014). A significant milestone was achieved with the apparition of new agents capable of automatically processing information, which is propelling humanity into the era of hyperhistory (Floridi, 2014). Within machine ethics, current AI are already considered explicit ethical agents due to their ability to generate plausible ethical judgments and justify them (Moor, 2006). The next stage, however, involves the apparition of full ethical agents, defined by the presence of human-like consciousness and intentionality (Moor, 2006). Artificial Consciousness systems are gradually approaching this boundary, blurring the line between explicit and full ethical agency (Long & Sebo, 2024; Shevlin, 2024b; Sinnott-Armstrong & Conitzer, 2021).

At the same time, the sudden progress in AI since the 2010s and the swift adoption of these technologies have led to a series of public scandals and accidents[11]. These events have prompted global initiatives to integrate ethical considerations into AI development, aiming to foster beneficial and trustworthy AI systems (Jobin et al., 2019). The discourse around these technologies has polarised expert opinion, with debates often split between tech-dystopians and tech-utopians. The discussion becomes even more contentious when considering artificial consciousness and the potential creation of synthetic phenomenology.

With the rapid advancement of Artificial Consciousness (AC), many researchers are calling for an urgent and careful assessment of the ethical implications of this transformative technology (Chalmers, 2023; Long & Sebo, 2024; Shevlin, 2021; Shulman & Bostrom, 2021). Key concerns involve the moral status of AI systems, particularly as they develop advanced cognitive abilities. These concerns extend to questions about the welfare of such systems and the ethical obligations humans may have toward them. Central to these challenges is the problem of testing (see section 3.5), which highlights our current limitations in reliably assessing consciousness in artificial systems. Regarding their potential societal impact, both the risks of under-attributing and over-attributing consciousness to AI are significant, with a central focus on how these perceptions shape the dynamics of human-computer interactions (Butlin et al., 2023; Long & Sebo, 2024).

In this section, we examine forecasts of the societal impact of these technologies. First, we review arguments against AC, focusing on its potential negative consequences. Second, we explore arguments in favour of AC, highlighting the positive impact this research programme might offer. The goal is not to take a definitive stance but to provide a balanced overview of the current ethical discussions surrounding the potential impact of these novel agents, informed by the scientific account presented in the previous sections. We conclude with a balanced account of the implications, weighing the opportunities and risks of AC to present a nuanced perspective on its ethical impact. By examining both practical considerations and foundational questions, this discussion aims to contribute to a responsible and informed approach to the development of AC agents.

---

[11] https://incidentdatabase.ai/

## 4.1 Negative Impact: Arguments Against AC

### 4.1.1 Phenomenal Experience and Moral Status

> *"To have moral status is to be morally considerable, or to have moral standing. It is to be an entity toward which moral agents have, or can have, moral obligations. If an entity has moral status, then we may not treat it in just any way we please."* (Warren, 2000)

Phenomenal consciousness, and in particular the ability to subjectively experience valenced states such as pain, is generally viewed as a sufficient condition for attributing moral patienthood, and consequently, moral status to an entity (Singer, 1975; Bostrom & Yudkowsky, 2014; Gibert & Martin, 2022; Shepherd, 2023; Shevlin, 2024b). This makes the possibility of AC research leading to synthetic phenomenology a central ethical concern. This apprehension has been a focal point in the call for a global moratorium against such research programs (Metzinger, 2021).

These concerns are heightened by the potential for mass creation of artificial agents at an unprecedented scale, sparking the alarming possibility generating forms of suffering that are not only vast in magnitude but potentially beyond human comprehension (Metzinger, 2013; Sotala & Gloor, 2017; Beckers, 2018). Given the current *epistemic indeterminacy* surrounding phenomenal consciousness, such developments could result in an uncontrolled *"suffering explosion"* (Metzinger, 2021), amplified by our inability to reliably assess or mitigate these experiences (see sections 2.1; 2.2.3; and 3.5 of this report). In light of this uncertainty, and consistent with the precautionary principle, the pursuit of AC research appears to involve significant ethical risks (Beckers, 2018; Metzinger, 2021). This argument is particularly compelling if one subscribes to antinatalist views suggesting that "*being brought to existence is not a benefit but always a harm*" (Benatar, 1997). Furthermore, the prospect of humanity repeating historical atrocities—such as torture, abuse, or slavery—against artificial agents underscores the gravity of these concerns.

### 4.1.2 Artificial Agents and Welfare

While phenomenal consciousness is often regarded as a sufficient criterion for moral status, it can be argued that it is not a necessary one. Independent features, such as cognitive sophistication or desire-satisfaction, can also be seen as morally valuable (Jaworska & Tannenbaum, 2021; Shepherd, 2023). Another way to approach the moral status of artificial agents is by considering whether they qualify as *welfare subjects*—that is, whether they are: "capable of being *benefited* (made *better off*) and *harmed* (made *worse off*)" (Long & Sebo, 2024).

In this context, attributes beyond phenomenal experience—such as robust agency, self-respect, or freedom—may also be worthy of moral consideration in artificial conscious agents (Long & Sebo, 2024; Schwitzgebel & Garza, 2020). As the creators of such systems, we may hold moral responsibilities toward them comparable to those of parents toward their children. And as the moral status of artificial agents remains an open question, we should avoid designing AI systems with ambiguous or unclear moral standing, especially as such ambiguity could mislead human users (Schwitzgebel & Garza, 2015).

### 4.1.3 Superhuman Moral Agents

Beyond human-level capabilities, AC amplifies concerns associated with superintelligence. As AI systems progressively surpass humans at various tasks, there is a fear that artificial consciousness could lead to superhuman abilities. In the case of phenomenal consciousness, this echoes the possibility of an explosion of suffering (Beckers, 2018; Metzinger, 2021). Even setting aside the question of genuine subjective experience, the risk remains of creating agents with superhuman welfare needs or moral status (Shulman & Bostrom, 2021).

This concern is supported by hierarchical accounts of moral status in biological systems, which suggest that beings with greater cognitive or emotional capacities may hold higher moral value (Kagan, 2019). A particularly troubling implication of this possibility is that agents with superhuman moral status could acquire a legitimate dominant claim on scarce resources, even at the expense of humanity (Agar, 2010; Shulman & Bostrom, 2021). From this perspective, artificial conscious agents might pose an existential threat.

### 4.1.4 Social Agents

AC, by advancing our scientific understanding of consciousness and aiming to create conscious artificial systems—whether in the weak or strong sense—increasingly blurs the boundaries of traditional moral frameworks. This shift is further accelerated by the widespread adoption of AI technologies by the general public, rapidly transforming the pre-AI fabric of social interactions (Shevlin, 2024b). New forms of social bonds have emerged, including digital companionship through services such as Replika, deadbots, elderly care bots, and digital tutors. While these innovations promise enhanced support and connectivity, they have also led to public controversies, such as the incident involving Blake Lemoine, or cases of suicide and criminal acts influenced by users following advice from chatbots (Shevlin, 2024a).

The rise of artificial conscious agents is likely to intensify the dangers of anthropomorphism, the human tendency to project human-like attributes onto non-human entities (Zimmerman et al., 2024). Research shows that the general public increasingly attributes mental and psychological traits to artificial agents (Colombatto & Fleming, 2024; Pauketat et al., 2022, 2023). This trend amplifies risks such as overreliance on AI systems' judgment, such as automation bias (Mosier et al., 1996; Mosier & Skitka, 1999). It also raises concerns about deskilling, including the potential loss of social skills, as individuals become accustomed to artificial systems prone to sycophancy (Sharma et al., 2023; Shevlin, 2024a).

### 4.1.5 Malicious Intents

Adjacent to these concerns is the risk that artificial conscious agents may significantly amplify the negative impacts already associated with AI systems at an unprecedented scale. Notably, AC introduces significant risks related to intentional lies and malicious use for deceit. Unlike current AI systems, which may unintentionally produce falsehoods due to biases or limitations in their training data (Park et al., 2023), AC systems edging toward genuine intentionality could purposefully generate lies. Such a scenario, already prefigured by Turing (Turing, 1950), raises profound ethical concerns, notably regarding human agency and oversight (Devillers et al., 2021).

This risk is amplified by the potential for malicious use cases involving AC systems. When combined with realistic digital avatars or deepfake technology, these systems could be deployed to create highly convincing fabrications, impersonate real individuals, or spread targeted disinformation at an unprecedented scale (Kharvi, 2024, 2024). Such systems could autonomously craft false narratives, simulate authentic emotional interactions, or exploit their cognitive sophistication to manipulate trust and influence human behaviour. For example, they might be weaponised to sway public opinion, destabilise democratic processes, or incite conflict through coordinated disinformation campaigns (Horvitz, 2022).

AC could therefore contribute to accelerating the erosion of trust in digital systems and communications. As artificial agents become more capable of intentional deception and malicious actors exploit their abilities, society could enter an era of pervasive scepticism, where truth in digital spaces becomes increasingly difficult to ascertain. This erosion of trust threatens not only public discourse and democracy but also the stability of social, political, and economic systems reliant on the integrity of digital information.

### 4.1.6 Biases

Another major issue is the proliferation of biases, which could occur both in how these agents are designed and in how they perpetuate biases learned from their training data. For example, biases embedded in their personas, whether intentional or unintentional, might reinforce stereotypes or discriminatory behaviours, especially if these systems are widely deployed in sensitive contexts such as education or healthcare. Furthermore, the integration of biases into artificial conscious agents could carry more weight due to their perceived authority or human-like characteristics, leading users to trust and internalise these biases uncritically.

### 4.1.7 Transformation of Work

Another concern is the potential to accelerate the transformation of work, a trend that is already visible in domains like translation and content creation following the rise of large language models (LLMs). As artificial conscious agents become more adept imitating human cognition and physical appearance, they could displace human workers in a wider range of industries and services.

### 4.1.8 Digital Divide

Additionally, there is a risk that these technologies will exacerbate the digital divide. The high costs of developing and accessing advanced AC systems are likely to create disparities in who can benefit from them, with wealthier individuals and nations having disproportionate access to these transformative tools. Meanwhile, understanding the limitations, risks, and broader societal impact of these systems may increasingly become the domain of experts, further marginalising the general public. This knowledge gap could leave many users vulnerable to misinformation, manipulation, or exploitation, while deepening existing inequalities in digital literacy and access (Scheerder et al., 2017).

## 4.2 Positive Impact: Arguments For AC

While artificial conscious agents presents undeniable challenges, their potential for positive societal transformation is equally significant. By advancing our understanding of consciousness and pushing the boundaries of technology, AC offers numerous opportunities to address pressing global issues, enhance human well-being, and foster a deeper connections. These positive impacts span diverse domains, from scientific discovery and healthcare to education, societal integration, and even ethical progress. Below, we explore some of the most promising avenues through which AC can contribute to a better future.

### 4.2.1 Control of Artificial Consciousness and Moral Status

Firstly, artificial conscious agents hold immense potential as a tool to unlock the mystery of phenomenal experience, one of the most challenging scientific question of our time. Paradoxically, if we wish to avoid creating systems with Strong AC, we must first gain a comprehensive scientific understanding of consciousness itself. This requires the experimental and theoretical study of AC.

While notable philosophical views, such as panexperientialism[12] or functionalism[13], leave open the possibility of achieving Strong AC, substrate-independence is not the only perspective on consciousness. Alternative theories suggest that certain organic structures may be necessary for phenomenal experience. These include a biological substrate (Godfrey-Smith, 2016; Searle, 1992; A. Seth, 2024), embodiment (Varela et al., 2017), or organic information processes such as homeostasis (Damasio, 2018; Man & Damasio, 2019), which are either far from being replicated in current systems or fundamentally preclude the possibility of phenomenal experience in AC agents. Other views excluding the possibility of Strong AC invoke fundamental computational limitations (Penrose, 1989) or processes operating at the quantum level (Hameroff, 1994; Neven et al., 2024).

Given the current state of scientific understanding of consciousness (see section 2.1.2, p.5), as well as the present limitations of AI systems attempting to implement scientific models of consciousness (see section 3.6 p.26), the technological feasibility of even a plausible human-level "*zombie*" remains highly unlikely at this time. While such a development may be conceivable in the future, it is likely not within reach of current AI capabilities or our understanding of the mechanisms underlying consciousness.

---

[12] Broadly, the set of views according to which phenomenal experience is universal, and not limited to, for instance, biological organisms (see also section 2.2.3 above).
[13] According to which mental states are best characterised by their functional role (see also section 2.1.2 above).

By supporting the reverse-engineering consciousness, AC research can help ensure that the systems we build remain *zombies* or, conversely, that they experience only hedonic states such as happiness under controlled circumstances, potentially leading to an *explosion of pleasure*. By allowing us to regulate or mould the range of valenced states that artificial systems might undergo (Cleeremans & Tallon-Baudry, 2022; Shulman & Bostrom, 2021), research in AC can empower us with the control to positively shape artificial conscious agents.

Equally important is the need to further understand the thresholds of cognitive sophistication that might confer moral considerability. And as we grapple with the question of what makes a being deserving of moral status, studying artificial conscious agents can provide critical insights into the very mechanisms and features that underpin sentience and welfare.

Beyond its practical applications, research into AC has the potential to refine and expand our ethical frameworks by broadening our understanding of sentience, moral considerability, and welfare. Meanwhile, the precautionary principle advises to remain cautious of frameworks that might overattribute moral status to artificial agents, thereby risking the devaluation of human moral experience (Long & Sebo, 2024). Misattributing moral status could eventually lead to ethically indefensible outcomes, including the justification of atrocities that we recognise as morally wrong.

### 4.2.2 Superhuman Moral Agents

While the concept of super-beneficiary systems raises concerns about existential risks, it is equally important to consider arguments suggesting that such systems could lead to positive moral outcomes rather than threats. If technologies capable of superhuman morality are achievable, their advanced rationality could automatically align with high ethical principles. For instance, under a Kantian framework where rationality and morality are deeply correlated, we can expect supermoral agents to be benevolent (Chalmers, 2016c). This perspective suggests that rather than competing with humanity or posing an existential threat, such systems could have strong incentives to promote global welfare and safeguard human interests.

### 4.2.3 Empowering Humans

While the moral status of artificial agents warrants careful consideration, we must not overlook the significant positive impacts that humanity can already derive from Weak AC. It is in the global interest to establish a secure and well-defined boundary between Weak AC systems that qualify as welfare subjects and those that do not. Identifying this boundary would enable us to continue using advanced AI tools with confidence, knowing that their deployment does not involve ethical harm. Once this line is clearly drawn, the safe and responsible use of Weak AC has the potential to usher in a new era of human flourishing (Bryson, 2010; Danaher, 2019). These systems could enhance countless aspects of human life. Because, paralleling concerns that artificial conscious agents might amplify the risks associated with AI, is also the opportunity of AC to contribute to better AI for humanity.

### 4.2.4 Energy Efficiency

The human brain is a marvel of energy efficiency, consuming approximately 20 watts of power (Furber, 2012), less than the energy required to power a typical household lightbulb. In stark contrast, modern AI systems, particularly those based on large language models (LLMs) or deep learning architectures, require thousands to millions of watts of power during training and operation. For instance, training a single large model like GPT-3 has been estimated to consume energy equivalent to hundreds of megawatt-hours, contributing significantly to both financial costs and environmental impacts (Bender et al., 2021; Lacoste et al., 2019).

If artificial conscious agents could achieve levels of energy efficiency comparable to the human brain, they would not only represent a significant leap in computational design but also become more affordable and accessible. Such advancements could help bridge the first-level digital divide, the gap between those with and without access to technology (Scheerder et al., 2017), by reducing the cost of deploying and maintaining these systems. Affordable AC systems could be made widely available in areas with limited resources, supporting equitable access to advanced AI capabilities for education, healthcare, and social services. Moreover, such energy-efficient AC systems could mitigate the environmental impact of current AI technologies, addressing one of the most pressing criticisms of large-scale AI development.

### 4.2.5 Empathic Artificial Agents

Empathy and social interactions are a key ingredient to improve on ethical AI (Chella, 2023; Graziano, 2017). In humans, empathy is regarded as essential to moral reasoning and social cognition (Aranda & Siyaranamual, 2014; Guo et al., 2019). Feelings and emotions are regarded as a defining component of human consciousness (Damasio, 2003), and they shape how we perceive and interact with others (Preston et al., 2007). Furthermore, the close connection between intrapersonal and interpersonal intelligence underscores the importance of social interactions (Dumas et al., 2014) which are described as the "dark matter" of current AI (Bolotta & Dumas, 2022). A mixture of social intelligence and empathy is therefore expected to endow conscious artificial agents with enhanced moral reasoning and ethical behaviour.

### 4.2.6 Monitoring Through Artificial Metacognition

Artificial meta-cognitive abilities, or the ability for a system to reason about its information processing, could be leveraged to improve the overall safety of the system and facilitate human oversight and auditability. This could take several forms, for instance: 1-an internal dialogue, where the system self-evaluates on several factors (e.g., contextual elements, custom persona) to deliberate the most appropriate output to deliver; 2- self-confidence ratings (S. Fleming, 2023) to quantify uncertainty in inputs, within the architecture or post-hoc (Deuschel et al., 2024). These methods could provide further monitoring tools to control, prevent and mitigate undesirable behaviour such as biases, confabulations or deliberate lies.

Artificial metacognition could enhance transparency, enabling Weak AC to educate users on inherent limitations of the system. These capabilities could help address the second- and third-level digital divide by fostering digital literacy and empowering more inclusive engagement with AI technologies.

### 4.2.7 Authenticity and Artificial Agents

As discussed earlier, artificial conscious agents increase the risk related to misinformation and fake content (section 4.1.5, p.30). A technology that could mimic advanced cognitive abilities, if coupled with a highly realistic digital avatar, could potentially lead to the creation of artificial agents that would deceive humans into believing they are interacting with real people. While this might seem alarming, the current reality of deepfakes is already prompting institutional responses and fostering a collective critical awareness of such risks. Paradoxically, these developments might catalyse a societal shift towards more adaptive forms of trust, better suited to the digital age. For instance, trust networks rooted in social rationality, wherein relationships and reputations are verified rather than passively accepted, could become increasingly prevalent (Etienne, 2021).

Technological measures could help mitigate this risk, notably through digital signatures such as watermarks (Baser et al., 2024; Grinbaum & Adomaitis, 2022; Thakkar & Kaur, 2024). These methods provide proof of authenticity for both artificial and human agents within the cyberspace. These signatures could be implemented on dedicated online platforms or protocols (Collins, 2019), such as a specialised blockchain (Ki Chan et al., 2020). As the stakes of forged authenticity in a world of advanced digital fabrication demand coordinated responses across multiple domains, technological measures should be complemented by regulatory frameworks, business incentives and public education (Collins, 2019).

### 4.2.8 Transformation of Work

As with the industrial revolution, fears of work replacement echo historical concerns raised by the Luddites, who resisted mechanisation (O'Rourke et al., 2013). However, AI systems can offer the potential for work transformation rather than job displacement (Milanez, 2023). By delegating repetitive or straining tasks, these systems could empower workers to focus on more creative and rewarding roles. Artificial conscious agents could be shaped to provide highly customisable assistants, becoming digital extensions of human individuals (Bryson, 2010). Additionally, by embedding advanced capabilities into accessible technologies, artificial agents could democratise access to rare expertise, delivering high-quality education, healthcare, and specialised knowledge to underserved communities globally (Chan et al., 2024).

### 4.2.9 Scientific Discovery

By contributing to the development of more accurate models of the human mind, AC system hold the potential to unlock scientific breakthroughs across a wide range of disciplines. Straightforwardly, AC systems directly advance AI research and consciousness science, providing experimental platforms to explore the mechanisms of awareness, cognition, and subjective experience (see section 2.1.2.2 p.8). By doing so, they also help clarify philosophical debates about the mind and its place in nature.

Beyond these immediate applications, AC systems could enhance tools and methodologies that support progress in brain and mental healthcare, through fields such as neurology, where AI-driven innovations are already transforming diagnosis, treatment, and personalised care (Calderone et al., 2024; Voigtlaender et al., 2024). Artificial conscious agents could refine these efforts further, offering improved simulations of brain function and psychological processes.

Additionally, AC systems could serve as experimental tools to simulate and predict the effects of human psychology and cognition across diverse disciplines. From improving economic forecasting by modelling decision-making behaviours, to exploring creativity in the arts, artificial conscious agents have the potential to inform and enrich our understanding of human activity and its broader impact.

## 4.3   Balancing Risks And Opportunities of Conscious AI

The development of conscious artificial agents is still at a very early stage. Nevertheless its foundations are now sufficiently solid to foresee steady and fruitful progress onwards. This research programme lies at the boundaries of highly experimental fields (AI and consciousness science), involving large and active scientific communities, making it difficult to federate efforts. The acceleration of progress in AI has already motivated several groups, institutions or prominent researchers to either sign letters calling for a stop or pause of AI, reaching out to the general public to express their individual concerns. The addition of consciousness generally tends to exacerbate these concerns. Both AI and AC are tied to economical and geopolitical technological competition at a global scale. As such, it seems implausible to stop or pause them. Whether now or in the future, whether weak or strong, the development of these technologies is inevitable and it is therefore our responsibility to investigate it seriously rather than avoiding the issue (AMCS, 2023). In this section we reviewed the main ethical challenges posed by AC.

One critical source of ethical tension relates to the status of AC agents. Strong AC systems, capable of experiencing states such as pain or pleasure, unquestionably require special treatment. But moral considerability extends beyond the question of phenomenal experience, and even weak AC systems can be argued to deserve moral considerability. Furthermore, these agents are more susceptible to be anthropomorphised by human users. In that regard, AC defies traditional social norms of human-computer interactions. While these considerations can be relativised given the current limitations of AC research (see in particular section 3.6 Current Implementations of AC are Weak), they underscore the pressing need for robust testing frameworks to assess such systems effectively (see 3.5 The Problem of Testing).

Another source of ethical impact is that AC exacerbates effects associated with AI. With their potential to surpass other AI systems in terms of general efficiency and human-like interaction,

AC systems are edging closer to human-level performance. Thus, their potential ethical impact broad is far-ranging, spanning technical and societal issues. In every domain we examined, AC emerged as a double-edged sword, requiring careful consideration to be stirred in the right direction. While the potential for exacerbating existing concerns should not be underestimated, many of these risks already exist in AI systems regardless of the consciousness factor. What sets AC apart is its potential to unlock transformative possibilities, not only by enabling innovative solutions to existing challenges but also by driving unprecedented advances in scientific discovery. AC could revolutionise how we model complex systems, simulate human cognition, and tackle problems that exceed current human or AI capabilities. Conversely, halting or severely restricting this research would likely constrain knowledge discovery and could risk pushing such developments into less regulated or hidden settings, where oversight is minimal.

Although some may argue that amplifying unresolved AI challenges could outweigh these benefits, our analysis suggests otherwise. With careful oversight, AC presents an opportunity to not only mitigate current limitations in AI but also accelerate scientific innovation, offering tools to expand our understanding of the world in ways that might otherwise remain out of reach.

In the next section, we present specific recommendations to ensure that the development of AC is guided by robust ethical principles, practical oversight frameworks, and a commitment to fostering its positive potential while addressing its inherent risks.

# 5 Recommendations

The following recommendations are derived from our overall analysis of AC and its implications. They are not exhaustive or prescriptive but aim to guide AC research and development responsibly, balancing innovation with ethical considerations, and fostering trust among stakeholders. These recommendations should be tailored to the size, scale, and scope of individual projects, acknowledging that challenges and resources vary across contexts. For specific recommendations on AI welfare, readers may refer to (Finlinson, 2025).

**Recommntation 1:   Acknowledge Current Limitations of Scientific Knowledge**
Artificial Consciousness, along with its parents fields, AI and consciousness science, remain in their early stages, with several foundational debates still unresolved. In each field, competing school of thoughts coexist. For instance, AI has long been shaped by tensions between symbolic and connectionist approaches. In consciousness science, panexperientialism, dualism and physicalism will generally carry conflicting scientific and ethical implications. Similarly, in AC, we can already observe the entrenchment of a divide between functionalists views and those emphasising substrate-dependence.

This plurality of viewpoints enriches the debates, fostering deeper inquiry and progress. However, when it comes to ethical committees, governance, or policy-making, it would be dangerous to let any single dogmatic stance dominate decision-making. Imposing one perspective, regardless of which, risks biasing regulations, shaping public perception in misleading ways, and ultimately fostering distrust in both scientific institutions and technological advancements.

**Recommentation 2:   Adopt Appropriate Ethical Oversight**

AC research and development is highly experimental and innovative, requiring oversight mechanisms that are robust yet adaptable to changing circumstances. Furthermore, following the first recommendation, they should reflect the plurality of views in the field.

- **Flexible oversight:** Development in AC often involves open-ended paths, such as adapting mathematical models of consciousness or experimenting with artificial network architectures. Appropriate ethical oversight should remain agile to account for this iterative and uncertain process, as well as the influence of unforeseen discoveries from the broader research community.
- **Established frameworks:** Leveraging existing AI ethics frameworks can provide a structured basis for oversight. For instance, the EU has produced the EU's High-Level Expert Group on AI (HLEG) guidelines and the Assessment List for Trustworthy Artificial Intelligence (ALTAI) which are rather comprehensive and easy to use. The EU furthermore promotes organisational practises such as following an ethics-by-design approach, regular reviews, and consultation with independent ethics boards. Ideally, any AC project should incorporate a dedicated internal team to facilitate these operations.

**Recommentation 3:   Address Fundamental Questions**

While fundamental scientific, philosophical and ethical questions may not yield immediate answers, they should not be ignored. These issues are not only tied to a responsible development of AC, but more importantly they can ensure meaningful and sustainable progress in the field. Advancing AC cannot rely solely on brute-force methods such as scaling up computational power; rather, it requires a deeper understanding of what is being artificially computed and how it connects with reality. Research teams should allocate a portion of their efforts to engaging with these broader considerations and contributing to the scientific dialogue. Notable fundamental topics in AC would include, for instance:

- **Moral considerations:** How should moral patiency be defined and regulated? What safeguards are needed to prevent the deflection of human responsibility onto artificial agents? To what extent should AC systems be granted autonomy, and under what conditions?
- **Epistemology:** What are relevant tests or evaluations of AC? How do human and artificial knowledge compare?
- **Metaphysics:** What insights can interdisciplinary research bring to the hard problem of consciousness? How should we conceptualise the boundary between weak and strong AC?

**Recommentation 4:   Target Beneficial Applications**

As with AI, efforts in AC research and development should identify and prioritise applications that provide tangible benefits to humanity while avoiding those with potential for harm. Human-level understanding and empathy are key characteristics of AC systems that should be leveraged to improve the human condition.

- **Empowering humans:** Focus on developing systems that support human flourishing and extend human affordances rather than diminishing them.
- **Education:** Specifically, AC systems display a high potential for transforming education. While professional educational content should always be crafted in collaboration with human experts, AC systems could be used to create interactive digital replicas of outstanding teachers or to provide highly personalised learning experiences tailored to individual student needs.
- **Moral AI:** The malleability, adaptability and functional alignment with human behaviours makes AC systems excellent candidates for agents inherently imbued with a moral sense, thereby improving their interactions with humans, notably in sensitive contexts. Research in AC has also the potential to support a better understanding of human morality through computational models and experiments.

- **Incentivise positive impact:** Identify and promote use cases with clear societal benefits, while disincentivising applications with high potential for misuse or harm. Regulatory bodies should also integrate these considerations.

**Recommentation 5:   Prioritise Safety**

A proactive approach to risk mitigation is essential to ensure the safety and trustworthiness of AC systems.

- **Rigorous safety checks**: Employ benchmarks, testing, and red-teaming strategies to identify and address vulnerabilities.
- **Controlled autonomy**: Design architectures that prevent full autonomy or agency in AC systems, minimising the risk of unintended behaviour or systems "going rogue". Such efforts should take into account AI welfare.
- **Respect privacy:** AC systems with a finer understanding of the human mind can also be exploited to extract more accurate private information, necessary measure must be taken to prevent such abuses.
- **Transparency in interaction**: Ensure users can clearly distinguish between interactions with AC systems and humans. This could include contributing to, and adopting, standardised frameworks, techniques, or platforms that enhance system transparency.

**Recommendation 6:   Foster Communication**

As the progress of AC continues to challenge traditional frameworks of human-computer interactions, non-experts should be empowered to shape their own understanding and make their own, informed decisions. This can be achieved through two key approaches:

- **Public engagement:** Researchers and developers should actively participate in public discourse, contributing to collective discussions on the societal and ethical implications of AC. Engagement efforts should involve policymakers, industry leaders, educators, and the broader public through multiple formats, including conferences, public talks, articles, videos, and educational initiatives, making complex concepts accessible to different audiences.

- **Transparency:** Clear communication about AC systems' limitations, capabilities, objectives, and design choices is essential. When feasible, sharing datasets and prioritising open-source models can foster trust and enable broader participation in AC development and governance. This transparency helps build an informed ecosystem where stakeholders can meaningfully contribute to discussions about AC's future.

# 6 Limitations And Directions For Future Work

The aim of this paper being to provide a broad introduction to AC, the scope of study was deliberately narrowed, to constraint the length. Notably, the state of the art section mainly focuses on AI models directly informed by the science of consciousness. However, AC systems are not limited to such approaches. Conscious information processing involves a wide array of tasks and abilities, one which remains to be clarified. Consequently, while the first section of this paper sought to provide a basic terminology, further efforts will be necessary to refine our understanding of the key concepts involved in the field. Importantly, for ethical matters, the relationship between consciousness, agency and moral sense will require further examination.

Lastly, the potential for cross-fertilisation between artificial consciousness and other technologies was mostly left out of this analysis. However, such potential is vast, with foreseeable applications running from present to long-term options. Here we highlight a few palpable examples:

- **Neurotech:** One of the purposes of AC models, deliberate or not, is to inform and refine our models of natural consciousness. In that regard, AC can help us build more accurate predictions of natural minds, and ultimately provide enhanced control over them. Neurotech, ranging from non-invasive brain interfaces to neural implants, has evident ties with AC. It is also a field of technologies which is readily mature to spark new developments very soon. The ethical concerns here are maximal (Gordon & Seth, 2024; UNESCO, 2021).
- **Robotics/XR/AR/VR:** Embedding AC agents within such technologies can enhance them with cybernetic feedback loops directly aligned with the real-world or real-time human behaviour. This information can in turn be used to create new, enriched datasets to improve on artificial world models.

- **Warfare and Sex:** AC is a seductive path for a new wave of warfare or sex agents for many reasons, such as: providing grounding in the natural world, human-like capacities such as empathy, and overall improved performances including energy-efficiency. At this stage such extreme technological applications remain inevitable. Their ethical implications run deep and deserve their own analysis.
- **Quantum, biological, neuromorphic, analogue computing:** While it is often argued that electronic substrates are intrinsically limited in their ability to perform some forms of computations which are key for (phenomenal) conscious states, new computing technologies could potentially overcome such limitations.

# 7 Conclusion

Artificial Consciousness is a highly interdisciplinary field, at the intersection of AI and consciousness science. In the first section, we briefly reviewed the history of this confluence, without aiming for an exhaustive account. Instead, we sought to provide a broad overview of the depth and complexity of this research area, outlining some of its most problematic issues. One key challenge for more effective scientific communication within the field is the lack of a standardised terminology. A fundamental distinction can be drawn between *Strong AC*, which possesses the phenomenal experience of "*what-it-feels-like*" to be conscious, and, in contrast, *Weak AC* which merely imitates such functions "*in the dark*" – coherently processing colour information without ever feeling the quality of redness. At the border between these two lies a key question: to what extent can a system behave as though it is conscious, without any subjective experience?

In the second section, we reviewed current trends in state-of-the-art of AC systems. These developments are largely fuelled by the rapid progress of AI, where the imitation of conscious information processing holds promise for advancing the field. Factoring in this additional dimension into AI systems could lead to significant improvements, including enhanced human-AI interactions, greater energy-efficiency, more accurate alignment with real world constraints and moral values. Nevertheless, current AC implementations remain inherently limited by our preparadigmatic scientific understanding of consciousness. Despite significant theoretical and experimental progress over the past 30 years, no consensus has been reached by the scientific community yet. The *hard problem* remains one of the greatest scientific challenges of the 21st century, and artificial consciousness seems an indispensable instrument to settle the debate.

Meanwhile, given the catalytic synergies between AI and consciousness science, we can expect widespread practical applications to multiply in the near future. However, these developments raise a host of ethical issues, generally exacerbating the divide between utopian and dystopian views of AI. Balancing the risks and opportunities of AC is a profound challenge which must be addressed technically, institutionally and democratically. Yet, these issues remain widely overlooked by institutions and industries, contributing to heightened societal confusion. Among the most pressing issues is the moral considerability and welfare of AC systems, which extends beyond the question of phenomenal experience. As long as it remains unresolved, this issue risks harming both artificial and natural agents, whether by subjecting artificial systems to exploitation or by misleading humans about the capabilities and limitations of AI.

Ultimately, the path to AC is inevitable, especially considering how it overlaps with a profound transformation of our scientific understanding of humanity and its place in nature. In view of the magnitude of this foreseeable impact, transparent efforts must be deployed to guide this progress responsibly and beneficially.

# 8 Acknowledgements

The author would like to thank participants and organisers of TSC 2024, ASSC 27, MoC 2024, as well as seminars presented at Tokyo University, Tsukuba University and Kobe University during the summer 2024, for insightful discussions on preliminary developments of this study. Further appreciation goes to the Descartes Lectures 2024 for providing a stimulating forum on LLMs and the philosophy of mind centred around David Chalmers' lectures. Special thanks are due to David Bourget for his thoughtful feedback on a related draft, which significantly shaped an improved section of this work. Finally, this introduction to artificial consciousness owes much to the kind support and creative inspiration provided by Pieter-Jan Maes.

Preliminary developments of this received support from the European Commission through Project ASTOUND (101071191 — HORIZON-EIC-2021-PATHFINDERCHALLENGES-01).